\documentclass[sigconf]{acmart}
\AtBeginDocument{%
  }
\usepackage{booktabs}
\usepackage{subcaption}
\setcopyright{acmlicensed}
\copyrightyear{2018}
\acmYear{2018}
\acmDOI{XXXXXXX.XXXXXXX}
\acmConference[Conference acronym 'XX]{Make sure to enter the correct
  conference title from your rights confirmation email}{June 03--05,
  2018}{Woodstock, NY}
\acmISBN{978-1-4503-XXXX-X/2018/06}

\begin{document}

\title{Rethinking Speech Codecs: From Compression to Autoregressive Generative Modeling}

\author{Yazheng Yang$^{\dagger}$, Yao Qiu$^{\ddagger}$, Hui Su$^{\ddagger}$, Qi Liu$^{\dagger}$}
\affiliation{%
  \institution{$\dagger$~The University of Hong Kong, $\ddagger$~Meituan Inc.}
  \country{China}
}
\begin{abstract}
Recent advances in speech language models leverage discrete speech representations from pretrained codecs to enable scalable training and generation. However, existing codecs are primarily optimized for compression without accounting for the autoregressive nature of language model training, resulting in suboptimal performance when modeling compressed speech tokens. In this work, we revisit speech discretization from a generative modeling perspective and propose a novel framework that explicitly aligns speech tokenization with autoregressive training. Our approach introduces autoregressive-compatible constraints during codec training, encouraging token sequences that exhibit temporal consistency and predictability. In addition, we propose a heterogeneous downsampling strategy for different layers of speech tokens, distinguishing semantic from acoustic layers, to improve the alignment between semantic tokens and corresponding textual content. Extensive experiments across multiple benchmarks demonstrate that our method bridges the gap between speech compression and generative modeling, enabling more effective continued pretraining of existing language models on speech data. The approach consistently improves performance across multiple codecs, validating its generality and applicability to diverse speech modeling scenarios.
\end{abstract}

\ccsdesc[300]{Computing methodologies~Speech recognition}
\ccsdesc[500]{Computing methodologies~Artificial intelligence}
\ccsdesc[200]{Information systems~Language models}
\ccsdesc[300]{Computing methodologies~Natural language processing}
\ccsdesc[300]{Computing methodologies~Machine learning algorithms}

\keywords{Speech Tokenization, Neural Speech Codec, Autoregressive Tokenization, Speech Compression, Speech Language Models}

\received{20 February 2007}
\received[revised]{12 March 2009}
\received[accepted]{5 June 2009}

\maketitle

\section{Introduction}
Large language models (LLMs) have achieved remarkable success across a wide range of natural language processing tasks~\cite{zhao2023survey,yang2024unleashing,yang2025qwen3}, and recent efforts have extended their capabilities to audio by leveraging discrete representations extracted from audio codecs~\cite{du2023lauragpt,xie2024mini,wu2024ts3,veluri2024beyond,defossez2024moshi}. These codecs, such as XCodec~\cite{ye2025xcodec}, typically employ Residual Vector Quantization (RVQ)~\cite{zeghidour2021soundstream,defossez2022high} with multiple codebooks, and encode the continuous waveform into multi-stream integer tokens, each representing a quantized latent sequence. At each time step, there are multiple integers, one per codebook, forming a richer representation. By mapping tokens back to their codebook embeddings and combining across multiple codebooks, one can recover the approximate latent representation of the original audio, which the decoder then converts back into waveform. The generated discrete tokens enable LLMs to process and generate audio using the same autoregressive modeling paradigm as in text.

\begin{figure}[htbp]
    \centering
    \includegraphics[width=1.0\columnwidth]{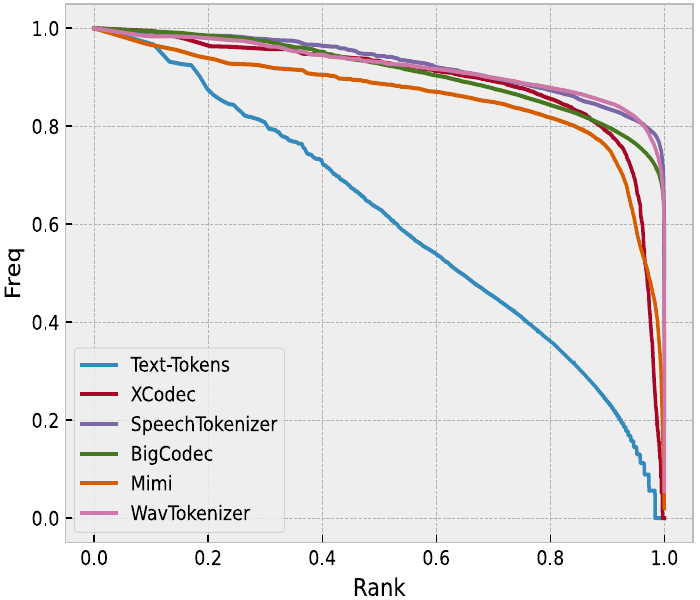}
    \caption{Zipf’s Law observed in various representative audio codecs, illustrating the power-law relationship between token frequency and rank commonly seen in natural language. The token distributions generated by existing representative codecs differ clearly from that of textual tokens, posing challenges for autoregressive modeling and generation.}
    \label{fig/intro_zipf_law}
\end{figure}

However, while such codecs are highly effective for compression efficiency and perceptual quality, their tokenization schemes are not explicitly designed with autoregressive modeling. This misalignment raises a critical issue: current audio tokenizers produce discrete sequences without modeling the conditional dependencies that autoregressive LLMs assume by design. As a result, token combinations are not necessarily consistent with the next-token prediction paradigm, weakening the temporal consistency and predictability of audio token streams. To further illustrate this discrepancy, we examine the statistical distributions of audio tokens generated by several representative codecs. As shown in Figure~\ref{fig/intro_zipf_law}, unlike textual tokens that naturally follow a Zipfian distribution~\cite{kingsley1932selected}, audio tokens deviate significantly from this power-law distribution. This divergence highlights the fact that current compression-oriented tokenizers do not capture the structured frequency-rank relationships that autoregressive models rely on, thereby increasing the learning burden during multimodal training, particularly when leveraging and continue training the pretrained LLMs. Addressing this gap forms a key motivation for our work.

In this work, we revisit speech tokenization from the perspective of generative modeling. We propose \textit{ARDDS} (\textbf{A}utoregressively \textbf{R}egularized training with \textbf{D}ifferent \textbf{D}ownsampling rates \textbf{S}trategy), a novel codec training framework that explicitly enforces autoregressive compatibility while aligning the temporal granularity of speech tokens with that of textual tokens. Specifically, our framework introduces an \textit{autoregressive regularizer}, which incorporates an auxiliary next-token prediction objective via an autoregressive decoder during codec training. This regularization aligns the structural properties of compressed speech tokens with the learning dynamics of large language models (LLMs), promoting token sequences that are not only compact but also autoregressively coherent. In contrast to conventional speech codecs that primarily optimize reconstruction fidelity, our approach explicitly accounts for downstream generative modeling requirements. 

In addition, we further reduce the frame rate of speech tokens to better match that of textual tokens, thereby narrowing the distributional gap between the two modalities. While recent speech codecs often adopt multi-layer token hierarchies to shorten token sequences, their overall token rate remains significantly higher than that of the corresponding text (e.g., XCodec's final token rate of 50 Hz), posing challenges for effective alignment with language models. The first layer of audio tokens, that typically capturing semantic rather than acoustic information, is particularly amenable to further compression. Since acoustic layers encode complex paralinguistic information, they require a higher temporal resolution. 
Motivated by this observation, we propose a \textit{heterogeneous downsampling strategy} that applies a lower sampling rate to the semantic token layer while preserving higher sampling rates for acoustic layers. This design effectively reduces the token rate without sacrificing essential semantic content. As a result, our method reduces the final frame rate to 6.25 Hz, substantially shortening token sequences and improving the efficiency of autoregressive modeling. By compressing the semantic token layer to a rate closer to that of the transcript, ARDDS achieves improved alignment between speech and text modalities. Consequently, the learned speech tokenization is better suited for integration with LLMs in cross-modal generative tasks, enabling more effective speech-language modeling.

Our approach is model-agnostic and can be integrated into various existing speech codecs. We demonstrate its generality by applying it to multiple representative codecs and validating its effectiveness through comprehensive experiments on speech generation tasks. Results show that our method not only maintains strong compression quality but also significantly enhances the performance of LLMs in speech modeling, paving the way for more effective large-scale speech language models (SpeechLMs) training. Our contributions are as follows:
\begin{itemize}
    \item We identify and systematically analyze the structural mismatch between conventional speech codecs and the autoregressive modeling paradigm employed by large language models.
    \item We propose an autoregressive-oriented codec training framework that explicitly encourages the generation of autoregressively coherent speech token sequences. In addition, our framework reduces the sampling rate of semantic tokens and aligns their temporal resolution closer to that of textual tokens, thereby improving computational efficiency and cross-modal compatibility.
    \item We demonstrate that our approach is general and codec-agnostic by applying it to multiple representative neural speech codecs. Extensive experiments show consistent improvements in generation quality, token statistical properties, and downstream large language model training and tasks, without degrading compression efficiency.
\end{itemize}

\section{Related Work}
\textbf{Audio Codecs and Discrete Representations}
Recent progress in neural audio codecs has enabled efficient compression of waveform signals into discrete token sequences~\cite{zhang2023speechtokenizer,defossez2024moshi,wu2024ts3,ye2025xcodec}. Models such as Encodec~\cite{defossez2022high}, XCodec~\cite{ye2025xcodec}, CodecBPE~\cite{shen2024acoustic}, and FunCodec~\cite{du2024funcodec} adopt vector quantization~\cite{zeghidour2021soundstream} and multi-level encoders to capture high-fidelity audio representations. These systems are typically optimized for rate-distortion trade-offs and perceptual metrics (e.g., PESQ~\cite{rix2001perceptual}, UTMOS~\cite{saeki2022utmos}), but are not designed to support the sequential dependencies required by autoregressive models. Although several codecs~\cite{kumar2023high,ye2025xcodec,yang2025almtokenizer} adopt causal architectures that prevent information leakage from future tokens into past hidden states, such architectures (e.g. XCodec~\cite{ye2025xcodec}) do not necessarily guarantee the autoregressive predictability of the generated speech tokens. This limitation is further evidenced by the statistical analysis presented in Figure~\ref{fig/intro_zipf_law}. More recent works like Moshi~\cite{defossez2024moshi} build on these representations for audio generative modeling, but majorly rely on multiple layer audio tokens organized with delay pattern. Our work diverges by modifying the codec training objective itself to make the resulting tokens inherently autoregressive-compatible.

\textbf{Speech-Language Modeling with LLMs}
Efforts to extend LLMs to audio include models like Llama-Omni~\cite{fang2024llama}, SpeechGPT~\cite{zhang2024speechgpt}, and SPIRIT-LM~\cite{nguyen2025spirit}, which typically process either continuous features or discrete tokens produced by codecs. While some models fine-tune pretrained encoders or align speech with text via contrastive learning (e.g., CLAP~\cite{elizalde2022clap}, BEATs~\cite{chen2022beats}), recent trends focus on representing audio as a language (e.g., GLM-4-Voice~\cite{zeng2024glm}) to reuse the autoregressive training paradigm within LLM through treating audio tokens as textual tokens. However, these approaches inherit the limitations of their underlying audio tokenizers, which are not optimized for next-token predictability. Our work addresses this limitation directly at the token generation level by regularizing the codec to produce LLM-adaptable token sequences.

\begin{figure*}
  \centering
  \centerline{\includegraphics[width=2.1\columnwidth]{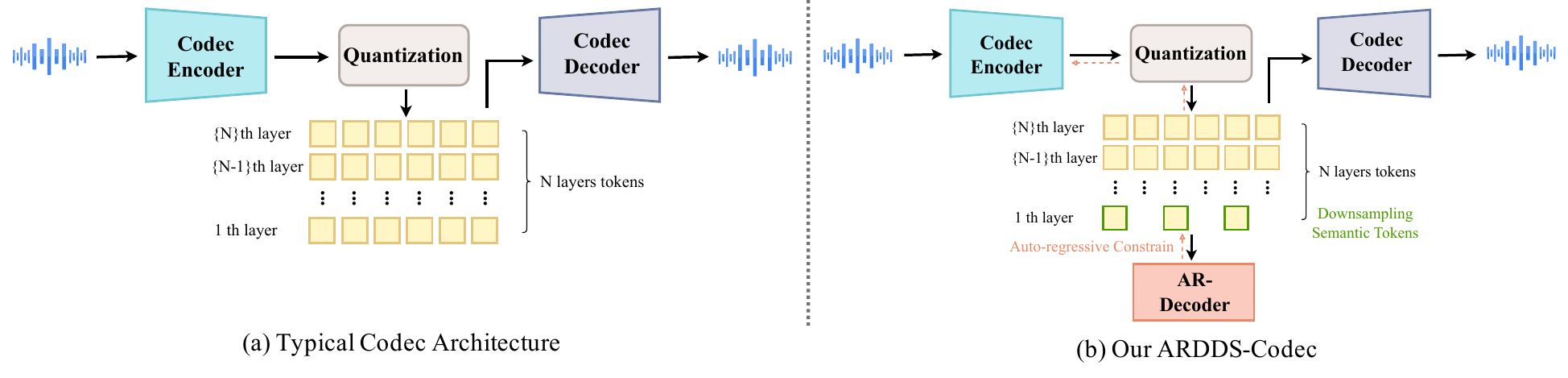}}
  \caption{\textbf{(a)} A general architecture of audio codecs, which convert raw audio into hierarchical token representations across multiple layers. \textbf{(b)} Overview of our method, that integrates standard audio codecs with autoregressive regularization and heterogeneous downsampling of the first-layer tokens. Downsampling the first-layer tokens brings the distribution of semantically rich speech tokens closer to that of textual tokens, while autoregressive regularization enforces the autoregressive prediction paradigm.}\label{fig/main_arc_fig}
\end{figure*}

Prior work in text highlights the importance of tokenizer alignment for generation quality, including token merging~\cite{wang2022deep,mao2025efficient}, vocabulary regularization~\cite{xu2021vocabulary}, and prefix tuning~\cite{li2021prefix}. In speech, approaches such as VQGAN~\cite{esser2021taming} and SpeechTokenizer~\cite{zhang2023speechtokenizer} explore expressive token spaces but largely overlook autoregressive compatibility. In contrast, we introduce an autoregressive decoder into the codec training loop, enforcing next-token predictability as an inductive bias to improve token consistency and bridge compression with generative modeling. Supporting evidence from vision includes LARP~\cite{wang2024larp}, which integrates an autoregressive prior for video tokenization.

\section{Methodology}
We propose a novel training framework for audio codecs that explicitly aligns the generated discrete token sequences with the autoregressive learning paradigm of large language models (LLMs), as shown in Figure~\ref{fig/main_arc_fig}. Our key idea is to incorporate an autoregressive regularization objective directly into the codec training process, ensuring that the resulting token sequences are both compressive and autoregressively predictable. This section presents the key components of our proposed methodology. We begin with a brief overview of typical audio codecs (§\ref{method/codec_intro}), followed by a description of our autoregressive regularization approach (§\ref{method/ar_intro}). Lastly, we introduce our heterogeneous downsampling strategy (§\ref{method/dds_intro}).

\subsection{Preliminaries}\label{method/codec_intro}

Let $x = [x_1, x_2, \ldots, x_T]$ denote a raw audio waveform, with $T$ being the total frame number. A conventional neural audio codec typically consists of three components: an encoder, a quantizer, and a decoder.

\textbf{Codec Encoder} The encoder $\mathcal{E}_\theta$ maps the continuous audio signal into a sequence of latent features:
\begin{equation}
z = \mathcal{E}_\theta(x) = [z_1, z_2, \ldots, z_N]
\end{equation}
where $z_i \in \mathbb{R}^d$ and the length of audio tokens $N \ll T$ due to downsampling.

\textbf{Quantization to Discrete Tokens} A central component in neural audio codecs is the quantization module $\mathcal{Q}$, which discretizes latent features into integers. The latent vectors $z$ are discretized by a vector quantization module (e.g., residual vector quantization (RVQ)~\cite{zeghidour2021soundstream,defossez2022high}), producing discrete token indices:
\begin{equation}
y = \mathcal{Q}(z) = [y_1, y_2, \ldots, y_N], \quad y_i \in \{1, 2, \ldots, K\}
\end{equation}
where $K$ is the size of the codebook. Depending on the codec, each $z_i$ may be quantized into one or more codebook levels. 
In addition, another quantization method known as Finite Scalar Quantization (FSQ)~\cite{julian2025finite} has been attracting increasing attention. In FSQ, each dimension of a latent vector is quantized independently into a fixed number of bins.

\textbf{Codec Decoder} The decoder $\mathcal{D}_\phi$ reconstructs the audio waveform from the discrete tokens:
\begin{equation}
\hat{x} = \mathcal{D}_\phi(y)
\end{equation}
The encoder $\mathcal{E}_\theta$ and decoder $\mathcal{D}_\phi$ are typically implemented using convolutional neural networks (CNNs) to exploit the local structure of audio signals efficiently. 

\textbf{Training Objective} The traditional codec is trained based on the GAN framework using multiple losses: 1) \textit{The reconstruction loss}: A multi-scale Mel-spectrogram reconstruction loss, computed as the L1 distance in the spectral domain across multiple scales. The Mel-spectrogram serves as a perceptually relevant representation of audio and is closely correlated with human auditory perception.
\begin{equation}
\mathcal{L}_{\text{recon}} = \| x - \hat{x} \|^2
\end{equation}
2) \textit{Least-square GAN loss}: Two types of discriminators are typically employed during codec training: Multi-Period Discriminator (MPD)~\cite{kong2020hifi} to capture pitch-dependent periodicity, and the Multi-Scale STFT Discriminator (MS-STFT)~\cite{defossez2022high}, which operates in the spectral domain to assess fidelity across resolutions.   
3) \textit{Discriminator Feature Loss}: Also known as perceptual loss~\cite{ledig2017photo}, this L1 feature-matching objective encourages the generator to produce perceptually natural outputs by aligning intermediate discriminator activations. 
4) \textit{Quantization Loss}: For vector quantization, we employ a quantization loss composed of a codebook loss and a commitment loss, which jointly align continuous encoder outputs with discrete codebook entries while stabilizing token usage. 

This standard training paradigm focuses solely on reconstruction quality and perceptual fidelity, with no constraints on the sequential structure of the token sequence $y$. As a result, the resulting tokens may not exhibit the kind of predictable, conditionally dependent patterns required by autoregressive language models.

\subsection{Autoregressive Regularization}\label{method/ar_intro}

To ensure that the token sequence $y$ is more compatible with autoregressive modeling, we introduce an auxiliary decoder $\mathcal{A}_\psi$ trained to predict the next token in the sequence, thereby encouraging the codec to produce token sequences that follow the autoregressive paradigm: 
\begin{equation}
\mathcal{L}_{\text{AR}} = - \sum_{t=1}^{N-1} \log P(y_{t+1} | y_{\leq t}; \psi)
\end{equation}
where $\mathcal{A}_\psi$ can be a lightweight Transformer-based decoder or RNN-based decoder. It is jointly optimized with the codec components.

A challenge in applying this autoregressive regularization is that discrete token IDs $y$ are non-differentiable, making it difficult to backpropagate gradients through the quantization module. To address this, we adopt a soft approximation of the discrete token assignment. Let $x_t^{\text{feat}}$ be the output feature from the encoder at timestep $t$, and let $\mathcal{C} = \{c_1, c_2, \dots, c_K\}$ denote the codebook. Instead of directly using the hard argmin operation to select the nearest codeword, we compute a soft assignment probability using a temperature-controlled softmax over negative distances:
\begin{equation}\label{eq_softmax}
p_t(i) = \frac{\exp\left(-\|x_t^{\text{feat}} - c_i\|^2 / \tau\right)}{\sum_{j=1}^K \exp\left(-\|x_t^{\text{feat}} - c_j\|^2 / \tau\right)}
\end{equation}
where $\tau$ is a temperature parameter. As $\tau \to 0$, $p_t(i)$ becomes a sharp distribution approximating one-hot, and the soft assignment converges to the argmin selection. This soft token distribution is then used as input to the autoregressive decoder $\mathcal{A}_\psi$, enabling gradient flow back through the quantization module and the encoder. 
The overall training objective is a combination of the original codec loss $\mathcal{L}_{\text{ori}}$ and the regularization:
\begin{equation}\label{eq_loss_com}
\mathcal{L} = \mathcal{L}_{\text{ori}} + \lambda \mathcal{L}_{\text{AR}}
\end{equation}
where $\lambda$ is a weighting coefficient that balances the trade-off between compression fidelity and autoregressive predictability. 
This joint training framework encourages the codec to produce token sequences that are both semantically meaningful for reconstruction and structurally aligned with next-token prediction, which is a critical property for continued pretraining in LLMs.

\subsection{Heterogeneous Downsampling}\label{method/dds_intro}
While audio codecs have adopted multi-layer token representations to reduce sequence length, a clear mismatch still remains between the number of speech tokens and the number of textual tokens of corresponding transcript. This discrepancy poses challenges when aligning audio with text in multimodal large language models. Considering a speech audio, there is potential to compress the semantic tokens to a length that more closely matches the number of text tokens in its transcript. Notably, the first layer tokens often contain content information that is more relevant to language modeling than acoustic layers, making them ideal candidates for further compression. 

To address this issue, we propose a \textit{heterogeneous downsampling} strategy that reduces the token rate of first quantization layer more aggressively than those of acoustic layers. Formally, for a codec producing $L$ token streams $\{y^{(0)}, y^{(1)}, \ldots, y^{(L-1)}\}$ from bottom (semantic) to top (acoustic), we assign downsampling rates $\{r_0, r_1, \ldots, r_{L-1}\}$ such that $r_0 < r_1=r_2= \ldots r_{L-1}$. This allows first layer tokens to appear at a coarser temporal resolution, aligning their frequency more closely with that of textual tokens, while preserving the finer granularity of acoustic details. Acoustic layers share the same sampling rate.
In our implementation, we apply downsampling at the feature level during quantization process. Specifically, for the first layer, we perform average pooling over the latent feature vectors within a fixed window size $W_{\text{ds}}$, effectively summarizing coarse information. This reduces the number of token emissions while maintaining representational fidelity.

By compressing the semantic layer to a lower token rate while preserving higher rates for acoustic layers, our design improves the alignment between speech semantics and textual representations. This results in speech token sequences that are better suited for autoregressive modeling in large language models (LLMs). 
Our approach is further motivated by recent Audio LLM frameworks~\cite{lelan2023stack,yang2023uniaudio,defossez2024moshi}, which adopt a delayed generation strategy: textual tokens are generated with a temporal lead of several steps, and the corresponding audio tokens are generated conditionally afterward, a paradigm commonly referred to as text-guided audio generation. This generation scheme has been shown to outperform purely speech generation, especially for long speech generation. Intuitively, when semantic speech tokens operate at a temporal resolution comparable to that of text tokens, they serve a role analogous to textual guidance, facilitating more coherent speech generation. 
Heterogeneous downsampling complements our autoregressive regularization (Sec.~\ref{method/ar_intro}), jointly encouraging the codec to produce speech tokens that are compact, expressive, and structurally aligned with the learning paradigm of LLMs. 
Our framework is modular and can be seamlessly integrated with a wide range of vector-quantization-based codecs, including XCodec, SpeechTokenizer, and BigCodec variants. This ensures high generalizability across different model families and application scenarios.

\section{Experiments \& Analysis}
We conduct comprehensive experiments to evaluate the effectiveness of our proposed autoregressive-compatible codec training framework. Our evaluation spans multiple codec architectures, statistical properties of generated token sequences, and end-to-end training of speech language models (SpeechLMs) following with downstream tasks' evaluation.

\begin{table*}
  \centering
  \scalebox{1.1} {
  \begin{tabular}{lcccc}
    \toprule
    Method / Task & StoryCloze\textuparrow & TopicStoryCloze\textuparrow & AIShell-I\textdownarrow & SeedTTS\textdownarrow \\
    \midrule
    XCodec~\cite{ye2025xcodec}          & 63.7 & 72.3 & 5.2 & 4.7 \\
    SpeechTokenizer~\cite{zhang2023speechtokenizer} & 63.4 & 72.5 & 4.9 & 4.6 \\
    XCodec-2~\cite{ye2025llasa-xcodec2}        & 64.2 & 72.8 & 4.7 & 4.2 \\
    BigCodec~\cite{xin2024bigcodec}        & 63.9 & 71.7 & 5.3 & 4.9 \\
    \hline
    XCodec:ARDDS(Ours)  & \textbf{70.1} \small{\textcolor{red}{+6.4}} & \textbf{76.9} \small{\textcolor{red}{+4.6}} & \textbf{2.5} \small{\textcolor{red}{-2.7}} & \textbf{2.3} \small{\textcolor{red}{-2.4}} \\
    \bottomrule
  \end{tabular}
  }
  \caption{Downstream performance of SpeechLM trained with different audio codecs. Evaluation metrics include accuracy (\%) for StoryCloze and TopicStoryCloze, character error rate (CER~$\downarrow$) for AISHELL-I (ASR), and word error rate (WER~$\downarrow$) for SeedTTS (TTS). These tasks span speech understanding and reasoning, speech recognition, and audio generation. The best results are highlighted in bold. Our proposed method consistently outperforms the baselines across different tasks.}
  \label{tb/speechllm_overall}
  \vspace{-2ex}
\end{table*}

\subsection{Experimental Settings}
\textbf{Codec}
For fair comparison, all codecs are trained on the same dataset composed of LibriSpeech and the Chinese split of Common Voice 23, with all audio resampled to 16 kHz. We evaluate our method on several representative codecs whose training implementations are publicly available. 
For each codec, we retain the original architecture and training hyperparameters, but standardize the training data and sampling rate across all experiments. For our autoregressive-compatible training, we set the softmax temperature in Eq.(\ref{eq_softmax}) to 0.01 after a grid search. This value results in a distribution that closely approximates one-hot behavior, without yielding further gains at lower temperatures in our experiments. The regularization weight $\lambda$ in Eq.(\ref{eq_loss_com}) is set to 1. 
We reduce the final frame rate of the first quantization layer to 6.25 Hz, while setting the remaining layers to 12.5 Hz, which shortens the overall sequence length. The grid search results about the final frame rate can be referred on the Appendix. 
Our autoregressive decoder is a lightweight Transformer decoder, with 6 layers, a hidden size of 2048, and 16 attention heads. It is jointly trained with the codec encoder and quantization module. We use XCodec~\cite{ye2025xcodec}\footnote{https://github.com/zhenye234/xcodec} as the representative multi-layer residual-vector-quantization (RVQ) codec to conduct our experiments.

\textbf{SpeechLM}\label{exp/ds_speechlm}
To evaluate downstream effectiveness, we perform large-scale SpeechLM pretraining by continuing training on a modified LLaMA-3~8B model. We introduce minimal architectural changes: (1) adding learnable embeddings for audio tokens and (2) using a separate LM head for audio. Speech tokens are embedded and concatenated with text tokens for autoregressive decoding. Following prior work~\cite{xie2024mini,defossez2024moshi,ye2025xcodec}, we adopt a delay-based generation strategy for multi-layer speech tokens. Training data consists of a large-scale Chinese speech corpus of approximately 400k hours, with all samples tokenized by the codecs under evaluation. Due to hierarchical sampling, codec layers operate at different frame rates. To align sequence lengths, we replicate each token in the first (semantic) layer across its corresponding time window to match the resolution of the other layers, forming a complete information at each position.

We design three training tasks: automatic speech recognition (ASR), text-to-speech (TTS), and interleaved text–audio modeling. The interleaving strategy segments long-form speech into blocks, where either audio tokens or the corresponding transcript tokens are randomly selected for each block. The resulting mixed-modality sequence is used as input to the SpeechLM, encouraging fine-grained cross-modal alignment and flexible generation. This approach has been adopted in prior work to facilitate audio–text alignment and generation~\cite{zhang2023speechgpt,defossez2024moshi,xie2024mini}. In practice, ASR and TTS are easier to optimize, while interleaved modeling better captures the complexity of cross-modal generation. Following GLM-4-Voice~\cite{zeng2024glm}, we use a dynamic sampling ratio of 90:1:1 for interleaved, ASR, and TTS tasks, respectively. The Audio LLM is trained using Megatron-LM\footnote{\url{https://github.com/NVIDIA/Megatron-LM}}
 on 32 NVIDIA A100 GPUs with the Adam optimizer ($\beta_1=0.9$, $\beta_2=0.95$, $\epsilon=10^{-8}$) and an initial learning rate of $1\times10^{-4}$. Training proceeds until convergence on held-out validation sets.

\subsection{SpeechLM Performance}
We first apply our method to XCodec and compare its performance with other open-source baselines. To examine its suitability for autoregressive modeling, we train speech language models on audio tokens generated by different codecs using the speech dataset described in §~\ref{exp/ds_speechlm}.

\textbf{Overall results} 
As shown in Table~\ref{tb/speechllm_overall}, the SpeechLM trained on tokens produced by \textit{XCodec:ARDDS} achieves substantial improvements across four downstream tasks, including two speech–language reasoning benchmarks (StoryCloze and TopicStoryCloze) and two generation-based speech tasks (AIShell-I for ASR and SeedTTS for TTS). These results consistently indicate that the audio tokens generated by our codec training framework are better suited for autoregressive modeling in SpeechLMs. The observed gains validate that enforcing autoregressive compatibility during codec training yields token sequences that better align with the next-token prediction dynamics of LLMs, thereby improving overall speech–language modeling performance.

\textbf{Impact to codec's original performance} A natural concern is whether the observed gains come at the cost of degraded codec quality. As shown in Table~\ref{tb/speechllm_xc_ablation} (right), the augmented codecs (“Ours”) achieve performance comparable to the vanilla baseline (“w/o both”) across standard evaluation metrics, including word error rate (WER) for reconstruction fidelity, speaker similarity (SPKSIM), perceptual naturalness (UTMOS), as well as objective speech quality and intelligibility measures (PESQ and STOI). This suggests that enforcing autoregressive compatibility does not compromise the codec’s original performance.

\begin{table*}
  \centering
  \scalebox{1.0} {
  \begin{tabular}{lcccc|ccccc}
    \toprule
    Method / Task & StoryCloze\textuparrow & TopicStoryCloze\textuparrow & AIShell-I\textdownarrow & SeedTTS\textdownarrow & WER\textdownarrow & SPKSIM\textuparrow & UTMOS\textuparrow & PESQ-nb\textuparrow & STOI\textuparrow \\
    \midrule
    Ours        & 70.1 & 76.9 & 2.5 & 2.3 & 4.13 & 0.88 & 4.30 & 3.53 & 0.95 \\
    \hline
    w/o AR      & 67.0 & 74.2 & 4.2 & 3.5 & 4.11 & 0.90 & 4.25 & 3.49 & 0.94 \\
    w/o DDS     & 67.1 & 74.7 & 3.8 & 3.4 & 4.14 & 0.90 & 4.27 & 3.51 & 0.95 \\
    w/o both    & 63.7 & 72.3 & 5.2 & 4.7 & 4.10 & 0.86 & 4.24 & 3.50 & 0.95 \\
    \bottomrule
  \end{tabular}
  }
  \caption{Ablation results of different copponents proposed in our method. The left part demonstrates the downstream performance of SpeechLM trained with different audio codecs, while the right part shows the analysis of codecs' original compression ability. The WER is conducted on the test set of LibriSpeech. The left part shows that both proposed components contribute to improved downstream performance, while the right part indicates that our method does not negatively affect the original tokenization performance of the codec.}
  \label{tb/speechllm_xc_ablation}
\end{table*}

\begin{figure*}
  \vspace{-3ex}
  \centering
  \centerline{\includegraphics[width=2.0\columnwidth]{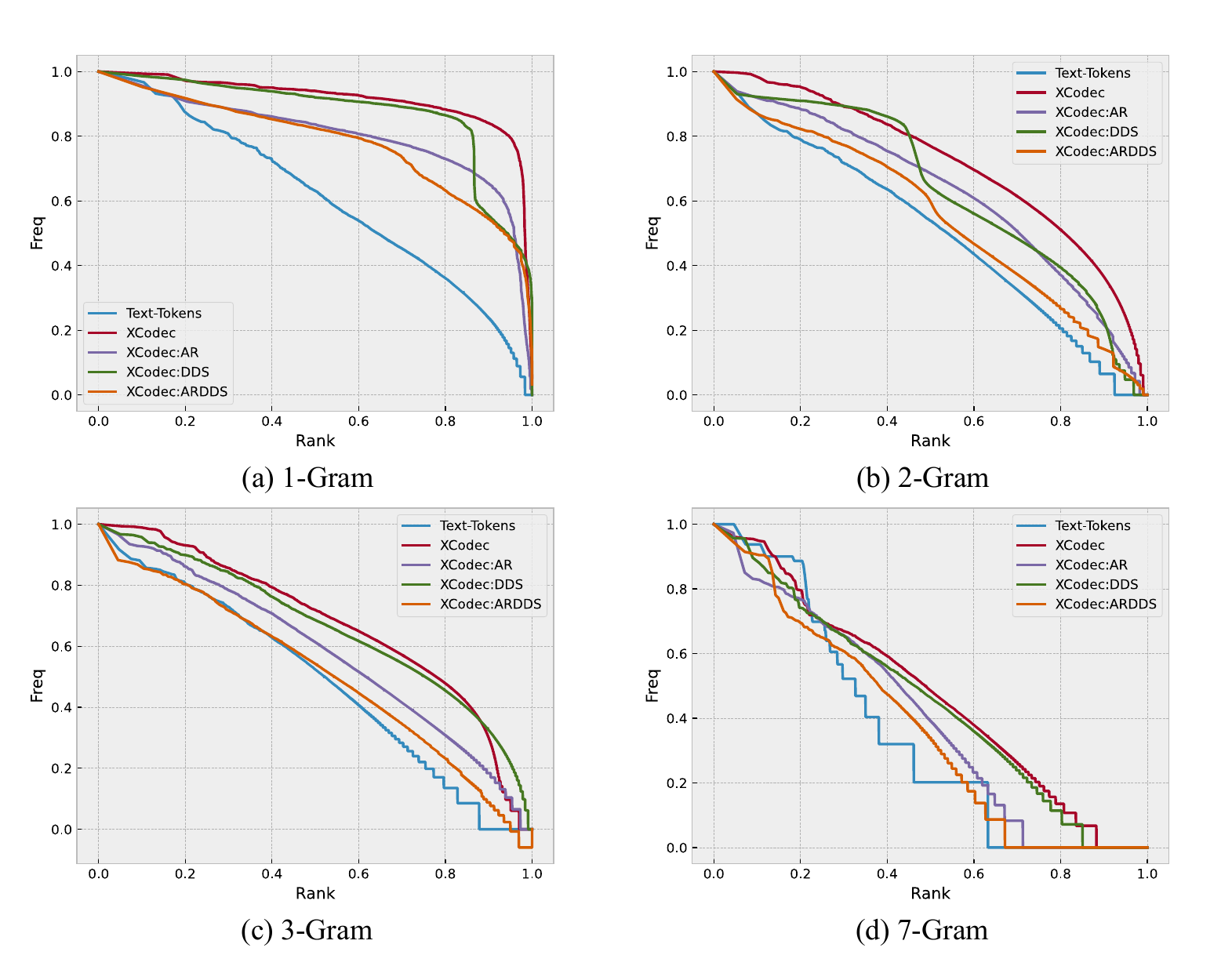}}
  \vspace{-2ex}
  \caption{Zipf’s Law analysis (normalized token log-frequency against normalized Log-Rank for several audio and textual languages) on 1-gram, 2-gram, 3-gram, and 7-gram token frequency distributions. ``XCodec:AR'' denotes XCodec with only autoregressive regularization, ``XCodec:DDS'' applies only heterogeneous downsampling strategy, and ``XCodec:ARDDS'' is our full method.}\label{fig/ab_zipf}
\end{figure*}

\subsection{Why Our Method Works} 

\textbf{The language of speech tokens}  
To further understand why our method improves autoregressive modeling, we analyze the statistical properties of the speech token sequences using Zipf’s Law. In natural language, token frequencies typically follow a power-law distribution, known as Zipf’s Law, where the frequency of a token is inversely proportional to its rank. This statistical regularity reflects the structured and compressible nature of language and has been widely used as an indicator of how well a token sequence is suited for autoregressive modeling and next-token prediction.

We adopt this perspective to examine whether audio tokens produced by different codecs exhibit similar Zipfian behavior. As illustrated in Figure~\ref{fig/ab_zipf}, token distributions from our proposed ARDDS-augmented method show a significantly more Zipf-like pattern across 1-gram to 7-gram statistics compared to baselines. This trend suggests that our autoregressive regularization effectively reshapes the token distribution to be more structurally aligned with natural language, thereby improving the predictability and learnability of the sequence for autoregressive LLMs.

\begin{figure}[htbp]
    \centering
    \includegraphics[width=1.0\columnwidth]{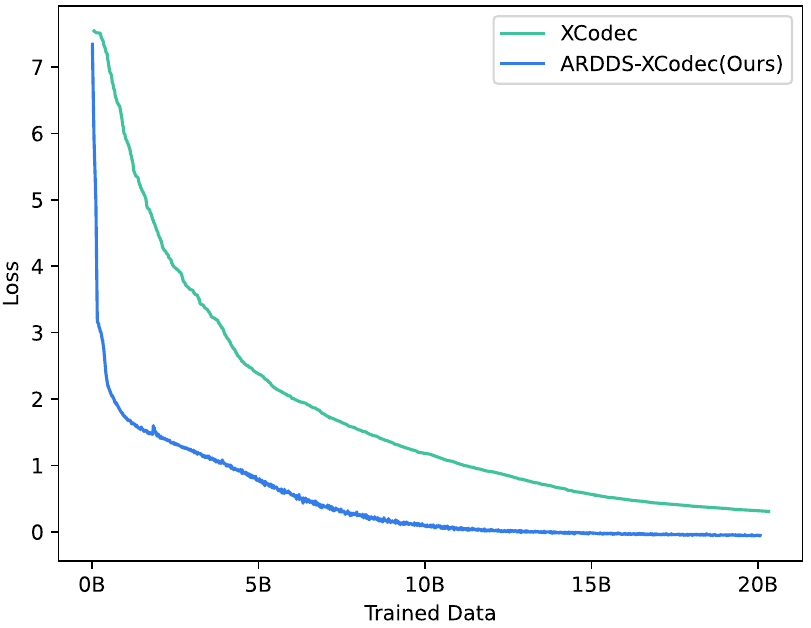}
    \caption{SpeechLM's training loss vs. trained data volume using XCodec and ARDDS-augmented XCodec. Our method leads to faster convergence and a lower final loss compared to the vanilla codec.}
    \label{fig/convergence}
\end{figure}

\textbf{SpeechLM Training Efficiency} As shown in Figure~\ref{fig/convergence}, SpeechLMs trained with speech tokens produced by the ARDDS-augmented tokenizer exhibit clearly faster convergence compared to those trained with the vanilla XCodec. The loss curve of our method drops more rapidly in the early stages and stabilizes at a lower final loss, indicating improved learning efficiency and better compatibility of the generated speech tokens with the learning pattern of existing LLMs.

\subsection{Ablation Study \& Analysis}
To better understand the contribution of each component in our proposed codec training framework, we conduct ablation studies based on the XCodec model. Specifically, we evaluate three variants: one without the autoregressive regularizer (w/o AR), one without the strategy of applying different downsampling rates (w/o DDS), and one with both components removed (w/o both). The results are shown in Table~\ref{tb/speechllm_xc_ablation}, which reports both downstream SpeechLMs performance (left) and intrinsic codec evaluation metrics (right).

\textbf{Ablation analysis on different components}
As shown in Table~\ref{tb/speechllm_xc_ablation} (left), removing either component results in consistent degradation across all downstream tasks, including reasoning (StoryCloze), speech understanding (AISHELL-I), and speech generation (SeedTTS). The full model (Ours) achieves the best overall performance, while removing both components leads to results of the original XCodec baseline. These findings confirm that the observed gains stem from the combined effect of autoregressive regularization and heterogeneous downsampling.

\begin{figure}[htbp]
    \centering
    \includegraphics[width=1.0\columnwidth]{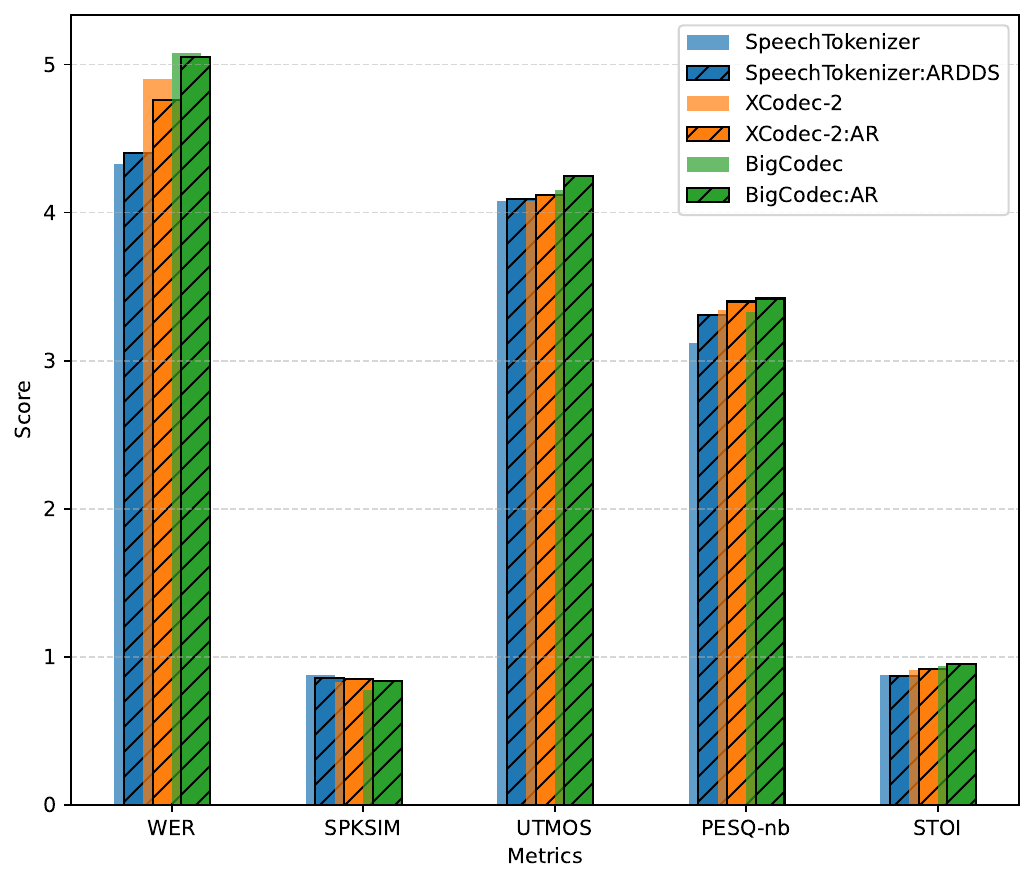}
    \caption{Performance comparison between the original codec and the corresponding codec augmented with our method. Our augmented codecs maintain performance without degradation compared to their vanilla counterparts.}
    \label{fig/other_codecs_ori_performs}
\end{figure}

\textbf{Impact on Codecs' original performance}
Importantly, none of the variants exhibits significant degradation in codec quality. Performance shows only subtle variations across configurations, indicating that our proposed modifications do not compromise the codec’s original capabilities.

\textbf{Zipfian's law analysis on different components}
To further support this observation, we examine the statistical structure of the generated token sequences. As shown in Figure~\ref{fig/ab_zipf}, which visualizes the $n$-gram token frequency distributions, our method (and its partial variants) consistently shifts the token statistics closer to the Zipfian distribution observed in natural language. The more Zipfian distribution of audio tokens indicates reduced redundancy and higher predictability, which in turn facilitates more efficient learning under the autoregressive modeling paradigm. These results suggest that our approach guides the codec toward generating more autoregressive-compatible token sequences, without degenerating compression quality.

This improvement can be attributed to the complementary effects of autoregressive regularization and heterogeneous downsampling. Specifically, autoregressive regularization constrains local token dependencies and encourages more structured sequential modeling, while heterogeneous downsampling reduces the effective sequence length by removing redundant temporal details in audio representations. This compression of the semantic token stream makes the resulting distributions closer to those of textual tokens, which are inherently more compact and information-dense. As a result, both components jointly reshape the frequency–rank behavior toward a more natural language-like distribution. Notably, this trend remains consistent from unigram to higher-order n-grams (up to 7-gram), suggesting that the improvements are not limited to local token statistics but extend to longer-range compositional structures. Overall, these results indicate that our method induces more linguistically natural token statistics, better matching the Zipfian properties of human language.

\textbf{Generalization Analysis}
To evaluate the generality of our autoregressive-compatible codec training framework, we apply it to three additional representative codecs: SpeechTokenizer~\cite{zhang2023speechtokenizer}, BigCodec~\cite{xin2024bigcodec}, and XCodec-2~\cite{ye2025llasa-xcodec2}. Among them, SpeechTokenizer and BigCodec are strong open-source baselines, while XCodec-2 is an upgraded version of XCodec that employs a single-layer finite scalar quantization (FSQ)~\cite{julian2025finite} scheme for improved simplicity and efficiency. For these codecs, we augment them with our autoregressive regularization (denoted as ''\texttt{XXX:ARDDS}'') while keeping the codec architecture and reconstruction objectives unchanged. Since XCodec-2 and BigCodec are single-layer models, we apply only the autoregressive regularization without heterogeneous downsampling (denoted as ''\texttt{XXX:AR}''). 
As shown in Figure~\ref{fig/other_codecs_ori_performs}, integrating our method does not degrade the original reconstruction quality or perceptual fidelity of the augmented codec. This suggests that the proposed autoregressive regularization operates orthogonally to the standard reconstruction objective, acting as a structural constraint rather than a competing optimization target. It reshapes the \emph{token distributional properties} rather than altering the codec’s information preservation and compression capacity.

\begin{figure}[t]
    \centering
    \includegraphics[width=1.0\columnwidth]{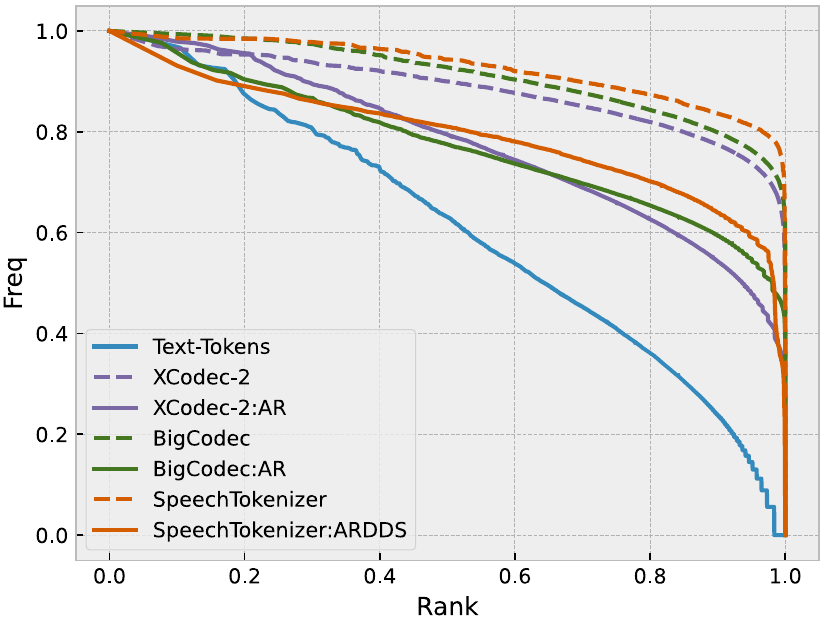}
    \caption{Plot of Zipf's Law comparison between baselines and our method augmented counterparts. Solid lines: ours; dashed lines: vanilla. Matching colors indicate the same baseline codec. Codecs augmented with our method exhibit a token distribution closer to that of textual tokens.
      }
    \label{fig/other_codecs_zipf_law}
\end{figure}

\begin{figure}[htbp]
  \centering
  \centerline{\includegraphics[width=1.0\columnwidth]{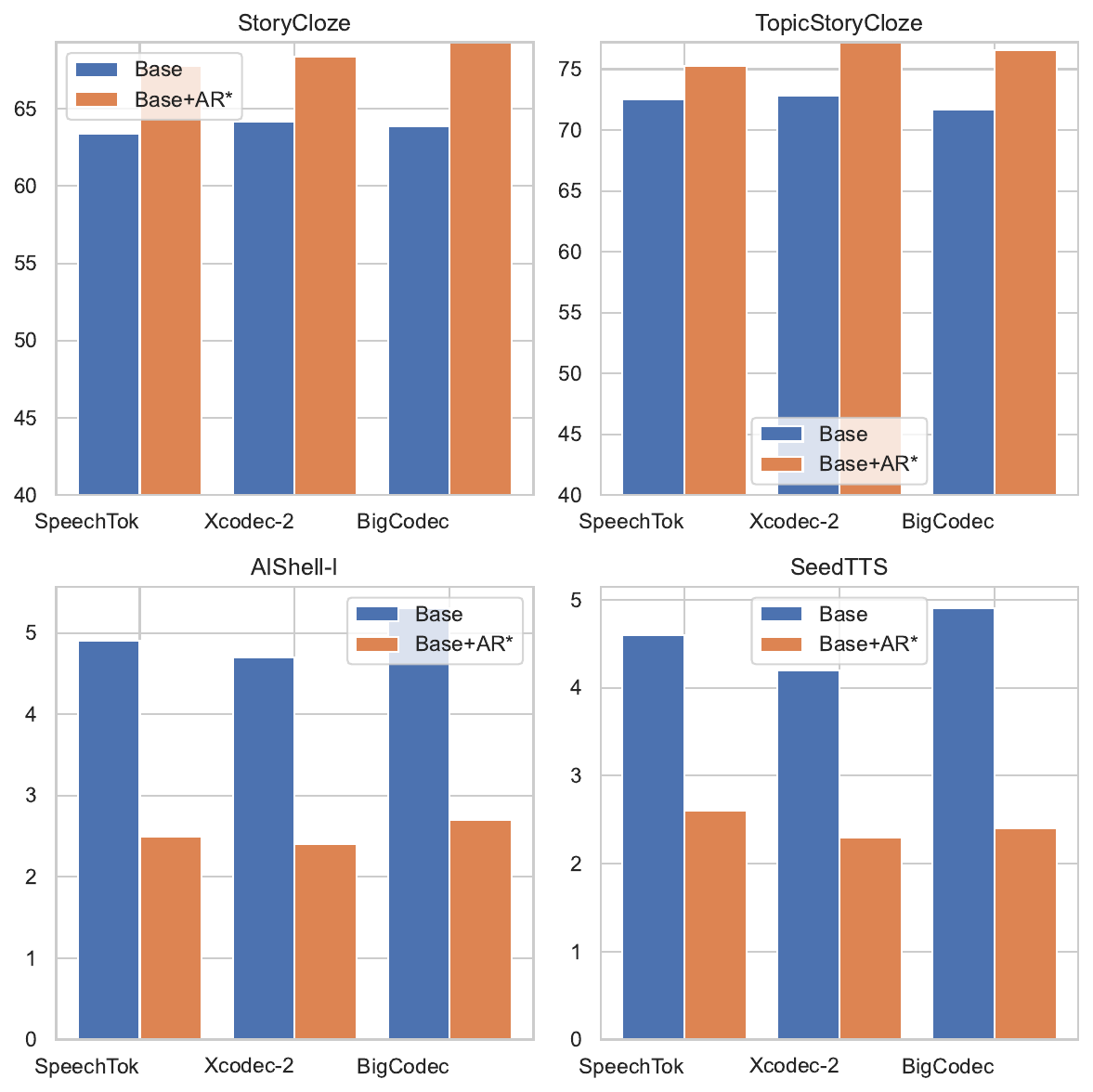}}
  \caption{Performance comparison between baseline models and their counterparts augmented with our method across multiple evaluation tasks for the trained SpeechLM, showing that our method consistently improves several existing codecs across different tasks.}\label{fig/other_codecs_speechllm_performs}
\end{figure}

We further analyze the statistical properties (1-gram frequency versus rank) of the generated speech token sequences through Zipf’s law, as illustrated in Figure~\ref{fig/other_codecs_zipf_law}. Across all codecs, the augmented versions consistently exhibit frequency–rank distributions that are closer to the linear trend observed in natural language. This shift is non-trivial: it indicates that autoregressive regularization does not merely improve local prediction consistency, but also induces a global reorganization of token usage statistics toward more natural language-like entropy distributions. From a modeling perspective, this phenomenon suggests that enforcing next-token prediction at the codec level biases the learned representation toward \emph{predictability-aligned compression}, where frequently co-occurring semantic units are assigned more stable and reusable token patterns. As a result, the long-tail structure of token usage becomes more pronounced and more aligned with Zipfian behavior, reflecting a more efficient allocation of representational capacity.

Finally, as shown in Figure~\ref{fig/other_codecs_speechllm_performs}, SpeechLMs trained on our augmented codecs consistently outperform those trained on their vanilla counterparts across a diverse set of downstream tasks, including StoryCloze, TopicStoryCloze, AISHELL-I, and SeedTTS. This consistent improvement across both understanding-style and generation-style tasks indicates that the benefits of our approach are not task-specific, but rather originate from a fundamental improvement in the tokenization space. 
Taken together, these results validate the strong generalizability of our framework across various codec architectures.

\section{Conclusion}
In this work, we revisit speech codec design from the perspective of autoregressive generative modeling. While existing codecs primarily optimize for compression fidelity, their discrete token outputs are not inherently aligned with the autoregressive learning paradigm required by large language models (LLMs). We introduce a general training framework that incorporates an autoregressive regularization objective and a heterogeneous downsampling strategy to encourage the generation of speech tokens that are both compact and autoregressively coherent. 
Our approach is codec-agnostic and applicable to a wide range of neural audio tokenizers. Through extensive experiments across multiple codecs, we demonstrate consistent improvements in the quality of audio token sequences for SpeechLM training. Further statistical analysis shows that our method induces more Zipf-like token distributions, aligning speech tokens more closely with natural language patterns. 
We hope this framework provides a practical pathway toward unifying compression and generative modeling in future audio modeling and generation systems.

\bibliographystyle{ACM-Reference-Format}
\bibliography{arcodec}

\appendix
\section{Minimum Final Token Rate Analysis}

\begin{figure}[htbp]
    \centering
    \vspace{-2ex}
    \includegraphics[width=1.0\columnwidth]{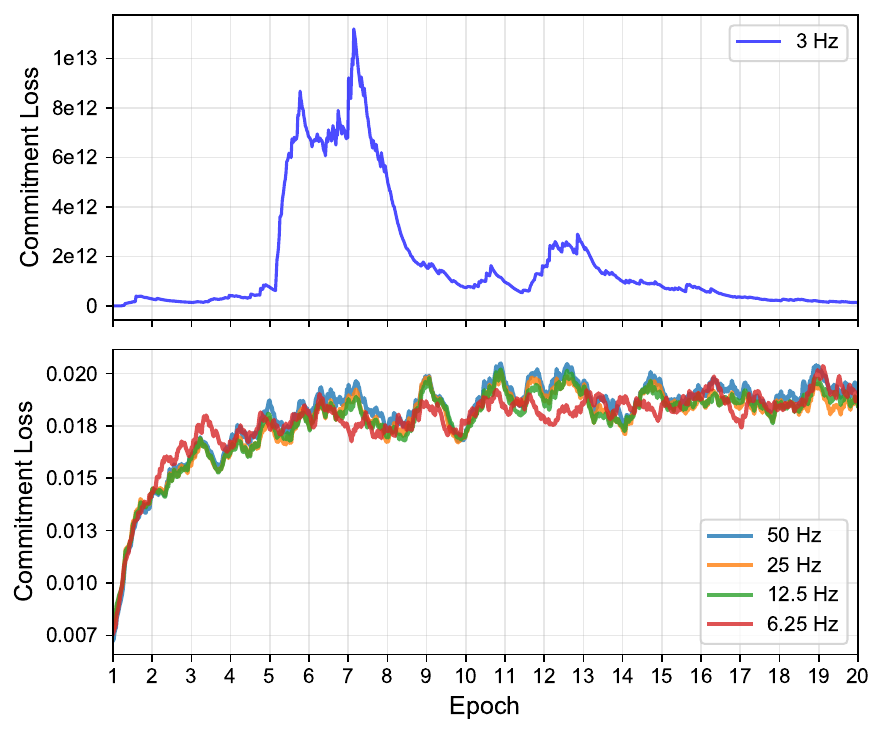}
    \vspace{-4ex}
    \caption{Training stability under different final frame rates. Commitment loss curves for codecs trained with final frame rates of 50 Hz, 25 Hz, 12.5 Hz, 6.25 Hz, and 3 Hz. Due to the substantially larger magnitude and instability of the 3 Hz setting, we present its curve in the upper subplot and the remaining configurations in the lower subplot. While training remains stable for frame rates down to 6.25 Hz, the 3 Hz configuration exhibits severe instability.}
    \label{fig/train_stability_different_frame_rates}
\end{figure}

To determine the lower bound of the final token rate that maintains stable and effective codec training, we conduct a controlled analysis by progressively reducing the final frame rate across multiple settings. This study aims to identify the minimal token rate at which the codec can still preserve semantic content while remaining trainable under an autoregressive learning paradigm. We vary the final frame rate of the codec across 50 Hz, 25 Hz, 12.5 Hz, 6.25 Hz, and 3 Hz. While all configurations down to 6.25 Hz exhibit stable training dynamics and comparable commitment loss trajectories, training becomes highly unstable at 3 Hz, as reflected by the large magnitude and high variance of the commitment loss (Figure~\ref{fig/train_stability_different_frame_rates}). The commitment loss measures how well encoder outputs commit to discrete codebook entries, and large or highly fluctuating values indicate unstable quantization and training dynamics. We therefore visualize the 3 Hz setting separately from the others to clearly highlight this instability.

Our results show that training remains stable down to 6.25 Hz, while further reduction to 3 Hz leads to severe instability, as evidenced by large and highly fluctuating commitment loss values. We attribute this instability to the insufficient number of speech tokens per second relative to textual tokens, which impairs the codec’s ability to represent semantic information effectively. These findings establish 6.25 Hz as the practical lower bound for the final token rate in our framework, and we adopt this setting in all subsequent experiments.

\section{Spectrogram Analysis of Reconstruction Quality}

\begin{figure}[htbp]
    \centering
    \vspace{-2ex}
    \includegraphics[width=1.0\columnwidth]{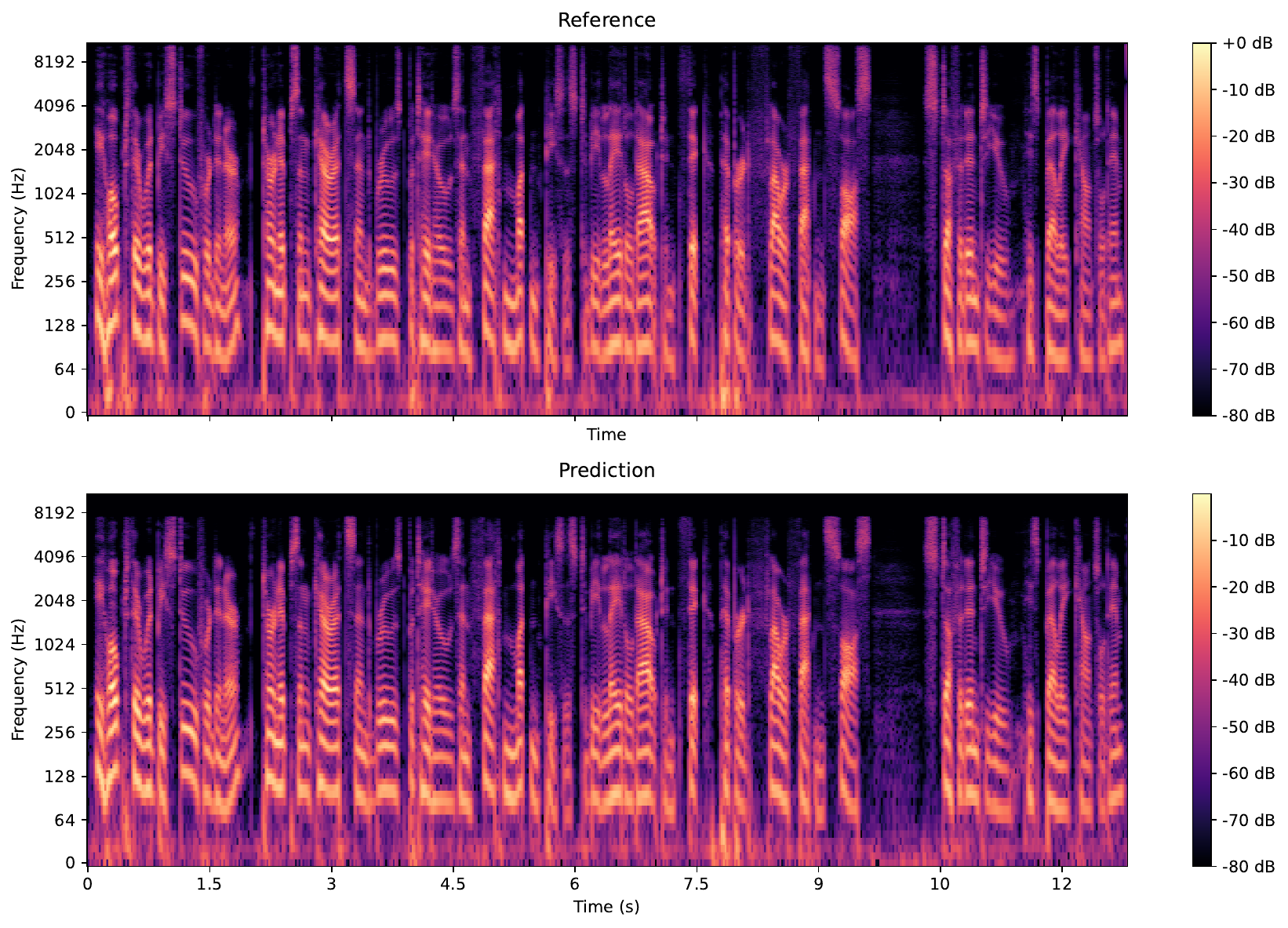}
    \vspace{-4ex}
    \caption{Spectrograms of reference speech and audio reconstructed by our codec (English sample from the LibriSpeech test set).}
    \label{fig/speech_visual_en_1}
\end{figure}

\begin{figure}[htbp]
    \centering
    \includegraphics[width=1.0\columnwidth]{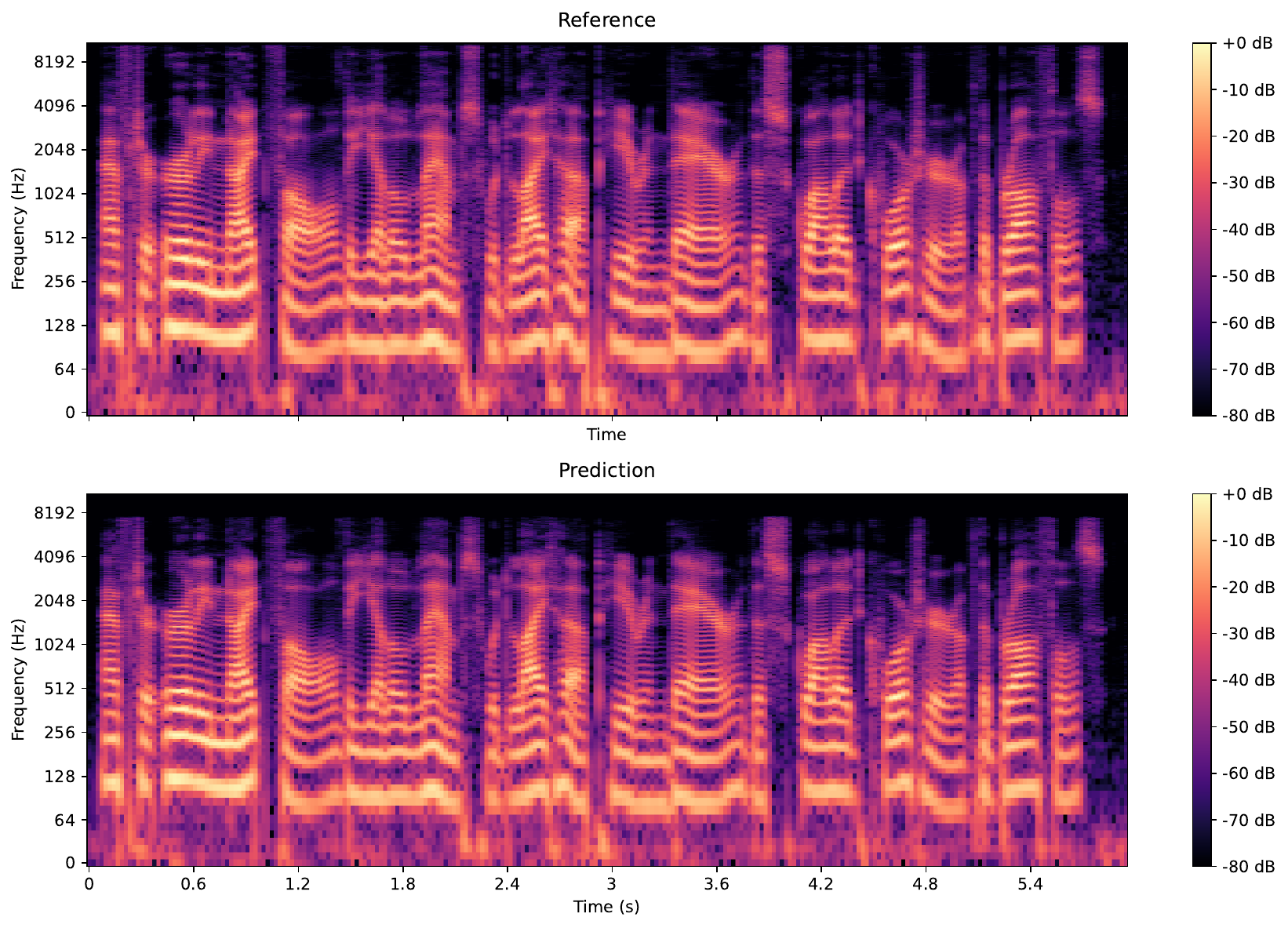}
    \vspace{-4ex}
    \caption{Spectrograms of reference speech and audio reconstructed by our codec (English sample from the LibriSpeech test set).}
    \label{fig/speech_visual_en_2}
\end{figure}

\begin{figure}[htbp]
    \centering
    \includegraphics[width=1.0\columnwidth]{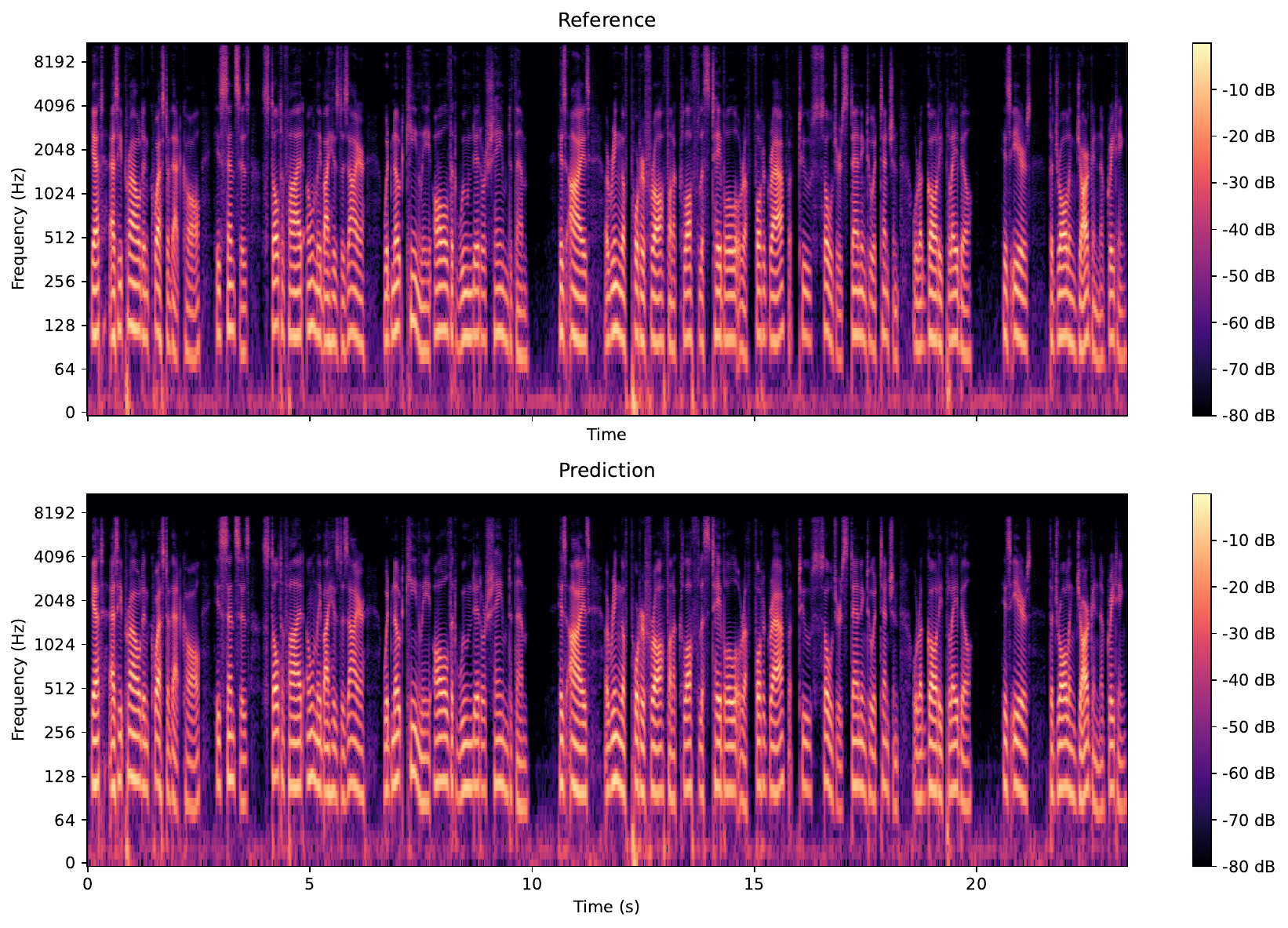}
    \vspace{-4ex}
    \caption{Spectrograms of reference speech and audio reconstructed by our codec (English sample from the LibriSpeech test set).}
    \label{fig/speech_visual_en_3}
\end{figure}

\begin{figure}[htbp]
    \centering
    \includegraphics[width=1.0\columnwidth]{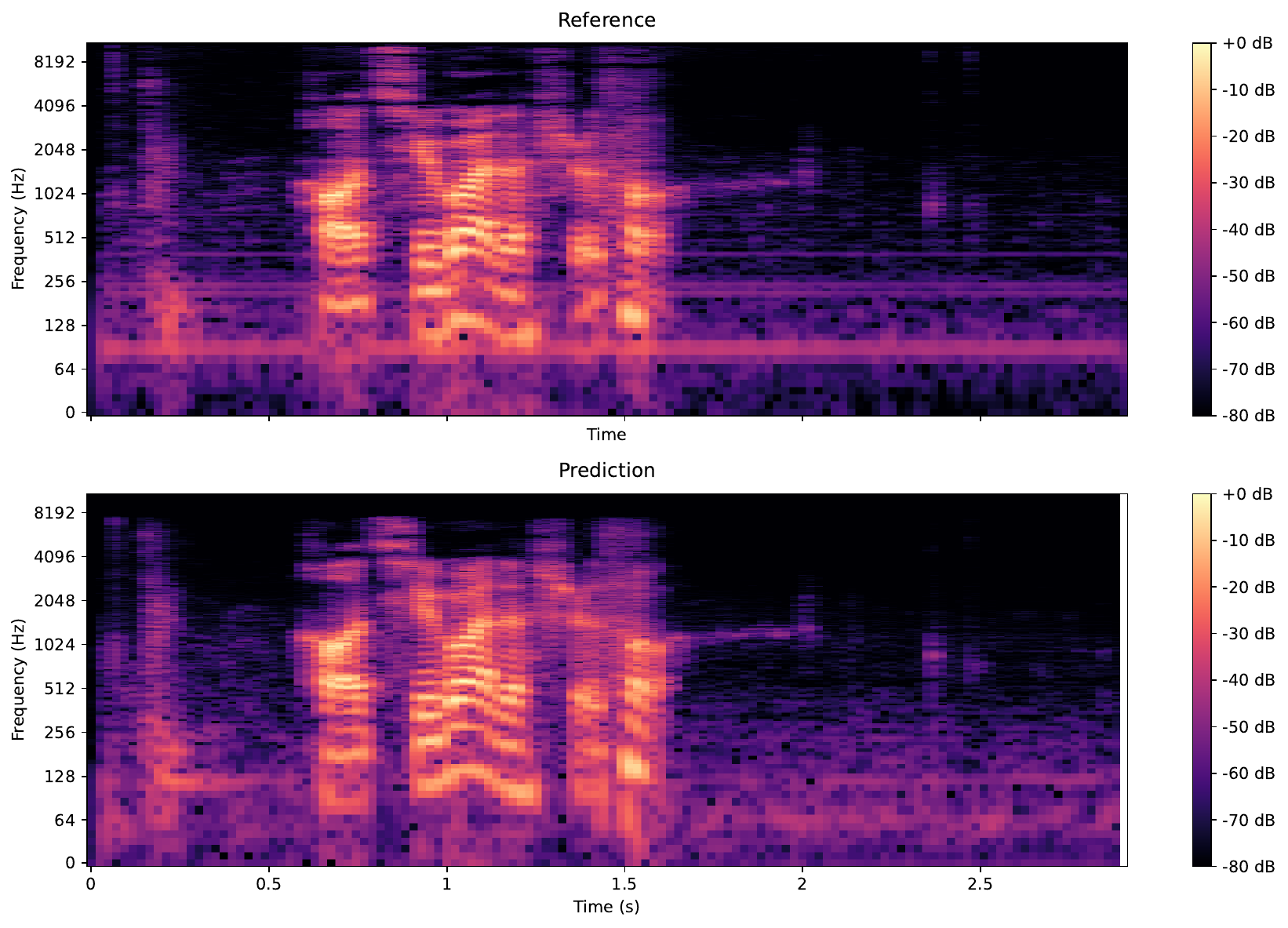}
    \vspace{-4ex}
    \caption{Spectrograms of reference speech and audio reconstructed by our codec (Chinese sample from the Common Voice test set).}
    \label{fig/speech_visual_zh_1}
\end{figure}

\begin{figure}[t]
    \centering
    \includegraphics[width=1.0\columnwidth]{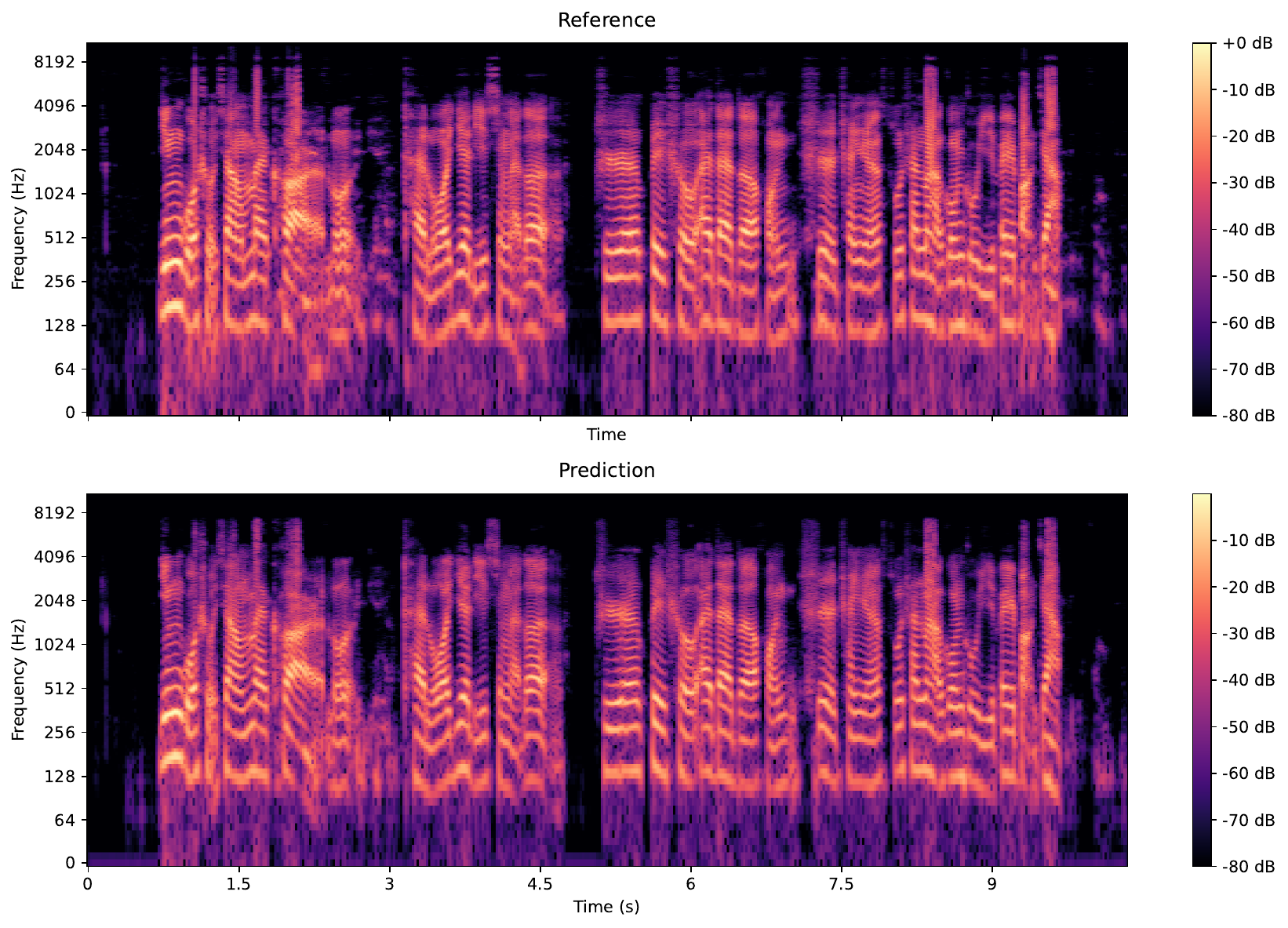}
    \vspace{-4ex}
    \caption{Spectrograms of reference speech and audio reconstructed by our codec (Chinese sample from the Common Voice test set).}
    \label{fig/speech_visual_zh_2}
\end{figure}

\begin{figure}[htbp]
    \centering
    \includegraphics[width=1.0\columnwidth]{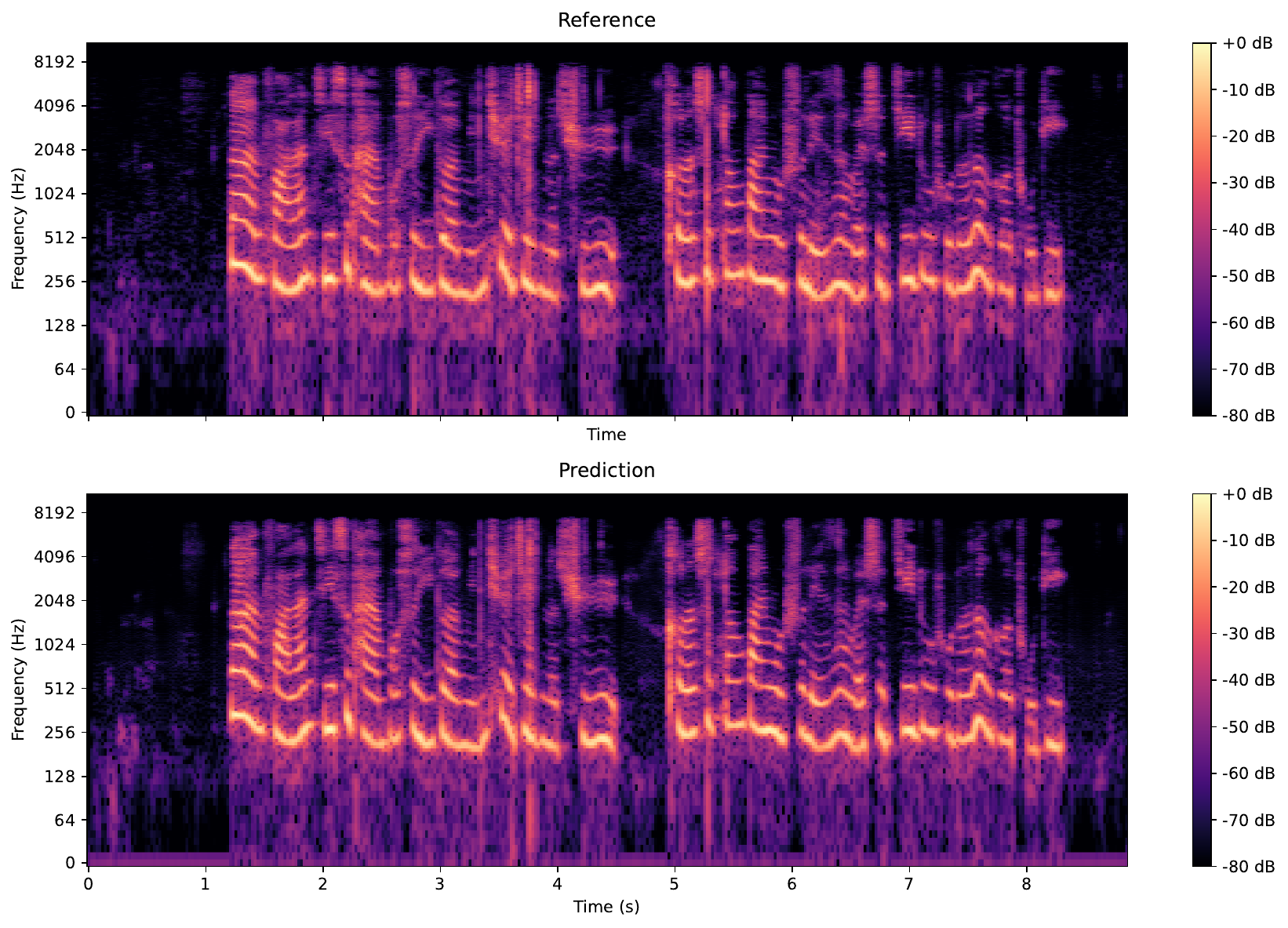}
    \vspace{-4ex}
    \caption{Spectrograms of reference speech and audio reconstructed by our codec (Chinese sample from the Common Voice test set).}
    \label{fig/speech_visual_zh_3}
\end{figure}

To qualitatively evaluate the reconstruction quality of our codec, we visualize spectrograms of reference speech and corresponding reconstructed audio across six test samples, including three English utterances from LibriSpeech and three Chinese utterances from Common Voice. As shown in Figure~\ref{fig/speech_visual_en_1}, Figure~\ref{fig/speech_visual_en_2}, Figure~\ref{fig/speech_visual_en_3}, Figure~\ref{fig/speech_visual_zh_1}, Figure~\ref{fig/speech_visual_zh_2} and Figure~\ref{fig/speech_visual_zh_3}, the reconstructed spectrograms produced by our method closely align with those of the reference speech, with no obvious degradation in time–frequency structure.

This visual consistency indicates that our training framework preserves both spectral detail and temporal coherence. Together with the objective metrics reported earlier, these results confirm that enforcing autoregressive compatibility does not compromise perceptual quality or reconstruction fidelity across languages.

\end{document}